\documentclass[twocolumn,showpacs,preprintnumbers,amsmath,amssymb]{revtex4}

\usepackage{graphicx}
\usepackage{dcolumn}
\usepackage{bm}
\usepackage{booktabs}
\usepackage{xcolor}

\begin{document}
\preprint{APS}
\title{Experimental characterization of the Gaussian state of squeezed light obtained via single-passage through an atomic vapor}
\author{P. Valente}
\author{A. Auyuanet}
\author{S. Barreiro}
\author{H. Failache}
\author{A. Lezama}

\email{alezama@fing.edu.uy} 
\affiliation{Instituto de F\'{\i}sica, Facultad de Ingenier\'{\i}a, Universidad de la Rep\'{u}blica,\\ J. Herrera y Reissig 565, 11300 Montevideo, Uruguay}
\date{\today}

\begin{abstract}
We show that the description of light in terms of Stokes operators in combination with the assumption of Gaussian statistics results in a dramatic simplification of the experimental study of fluctuations in the light transmitted through an atomic vapor: no local oscillator is required, the detected quadrature is easily selected by a wave-plate angle and the complete noise ellipsis reconstruction is obtained via matrix diagonalization. We provide empirical support for the assumption of Gaussian statistics in quasi-resonant light transmitted through an $^{87}$Rb vapor cell and we illustrate the suggested approach by studying the evolution of the fluctuation ellipsis as a function of laser detuning. Applying the method to two light beams obtained by parting squeezed light in a beamsplitter, we have measured entanglement and quantum Gaussian discord. 

\end{abstract}

\pacs{42.50.Lc,42.50.Dv,42.50.Nn} 

\maketitle
\section{\label{Introduccion}Introduction}

It is well established that the propagation of linearly polarized continuous-wave laser light through a near resonant atomic vapor can result in squeezing of the field with polarization component orthogonal to the incident field polarization (vacuum squeezing) \cite{MATSKO02,RIES03,MIKHAILOV08,MIKHAILOV09,AGHA10,BARREIRO11}. Such phenomenon has been understood as the consequence of the nonlinear response of the atomic medium giving rise to the effect of polarization self-rotation (PSR) of elliptically polarized light. PSR is the consequence of the third order susceptibility of the atomic sample. Using this susceptibility, it can be shown that the input-output relations for the quantum field operators \cite{RIES03} result in vacuum squeezing of the field polarized in the direction orthogonal to the incident field polarization (hereafter referred as ``vacuum polarization"). 

Initial work on this subject has centred the attention only on the squeezed polarization component of the field. From the experimental point of view, this required an auxiliary local oscillator to perform the balanced homodyne detection of the squeezed light. Usually, the strong linear polarization component of the field has been used as a local oscillator \cite{RIES03,MIKHAILOV08,AGHA10}.  

In a previous article \cite{BARREIRO11}, we have adopted a different point of 
view in which the two polarization components of the light beam are 
simultaneously considered. From this perspective, the squeezing effect is seen 
as a modification in the fluctuations of the polarization of the light i.e. the 
Stokes operators associated to the field. A reduction of the variance of a 
Stokes operator below the corresponding standard quantum limit (SQL) indicates 
polarization squeezing \cite{KOROLKOVA02}. Experimentally, the main advantage of using polarization 
modes for the investigation of the quantum properties of light is the 
elimination of the requirement of a local oscillator.

Polarization squeezing and entanglement were first 
obtained using two independent optical parametric amplifiers 
\cite{Bowen02a,Bowen02b} and a cold atomic cesium cloud  
 inside an optical cavity \cite{Josse03,Josse04a,Josse04b}. 
Taking advantage of the fact that polarization squeezing does not require a shared local oscillator, it was recently used for 
entanglement distribution \cite{Peuntinger13} and long distance 
quantum communication \cite{Peuntinger14}.

In this article we go a step forward in the examination of the fluctuation properties of the polarization state of laser light transmitted through an atomic sample by incorporating the additional assumption of the Gaussian statistics of the quantum state of the light field. These states have a very elegant mathematical description and are simple to manipulate. Also, Gaussian states of light are commonly encountered: the vacuum and the coherent states are Gaussian and so are the squeezed vacuum and the displaced squeezed vacua.  In addition, a wide range of field evolutions are known to preserve the Gaussian nature of the field state \cite{BRAUNSTEIN05}. In particular, the PSR effect results in squeezed vacuum of the vacuum polarization component. Based on the above, we have made the working assumption of the Gaussian nature of the light state. We provide experimental evidence in support of this assumption. 

The purpose of this article is to show that by combining the description of the field in terms of Stokes operators and the hypothesis of Gaussian statistics one can  easily have access to full information about the light state. We illustrate this possibility by studying laser frequency dependence of the squeezed and anti-squeezed quadrature variances and the corresponding quadrature angles. In the last part of the paper, we apply Gaussian analysis of Stokes operators measurements to the determination of entanglement quantifiers and quantum Gaussian discord on two light beams separated in a beamsplitter. 

\section{\label{Background}Background}
\subsection{Stokes operators}
A quasi-monochromatic light beam propagating in the $z$ direction is associated to the annihilation operators $a_x$ and $a_y$ corresponding to the $x$ and $y$ linear polarization components respectively. Alternatively, one can refer to a different polarization basis and introduce $a_{1(2)}=(a_x+(-)a_y)/\sqrt{2}$ corresponding to the polarization components along the main diagonals of the $x,y$ plane or $a_\pm=(a_x\pm ia_y)/\sqrt{2}$ corresponding to the two circular polarizations components. 

In order to describe the polarization state of the field it is convenient to introduce the Stokes operators:
\begin{subequations}\label{SS}
\begin{eqnarray}
S_{0}& =&  a_{x}^{\dagger}a_{x}+a_{y}^{\dagger }a_{y} =n_x+n_y \label{S0}\\
S_{1}& =& a_{x}^{\dagger}a_{x}-a_{y}^{\dagger }a_{y} =n_x-n_y \label{S1}   \\
S_{2}& =& a_{x}^{\dagger }a_{y}+a_{y}^{\dagger }a_{x} =n_1-n_2 \label{S2}  \\
S_{3}& =& i\left( a_{y}^{\dagger }a_{x}-a_{x}^{\dagger }a_{y}\right) = n_--n_+\label{S3}  
\end{eqnarray}
\end{subequations}
where $n_\nu\equiv  a_{\nu}^{\dagger}a_{\nu}$ is the photon number operator associated to a given polarization. The Stokes operators $S_{1-3}$ obey the angular-momentum-like commutation rule:
\begin{equation}
\left[ S_{i},S_{j}\right] =2i\epsilon _{ijk}S_{k},
\end{equation}
where $\epsilon _{ijk}$ is $\pm 1$ depending on the parity of the $i,j,k$ permutation. 

It is convenient to introduce a generalized Stokes operator in the $2-3$ plane parametrized by the angle $\theta$: 
\begin{eqnarray}
 S_\theta&=& n_{\theta +}-n_{\theta -} = S_2 cos(\theta)+S_3 sin(\theta) \label{giro} 
\end{eqnarray}
where $n_{\theta \pm}$ are the photon number operators corresponding to the field operators: $a_{\theta\pm}=(a_x\pm e^{i\theta}a_y)/\sqrt{2}$ which correspond to two orthogonal elliptical polarization components. It can be easily shown that $S_{\theta=0} = S_2$ and $S_{\theta=\pi/2} = S_3$. The generalized Stokes operators verify the commutation rule:
\begin{equation}
\left[ S_\theta,S_{ \theta^{\prime} }\right] = 2i S_1 sin\left( \theta^{\prime}-\theta\right)
\end{equation}

In this article we are concerned with a field that is intense and essentially polarized along the $x$ direction. Let $\alpha=\left\langle a_{x}\right\rangle $ be the expectation value of the field operator for the $x$ polarization. We assume that $\vert\alpha\vert \gg \vert \left\langle a_{y}\right\rangle\vert \simeq 0$ and that the fluctuations of the $x$ polarization components of the field are small compared to $\alpha$. Under such conditions the Stokes operators can be approximated as:
\begin{subequations}\label{SSapp}
\begin{eqnarray}
S_{0}& \simeq& S_1 \simeq \vert \alpha \vert^{2}  \\
S_{2}& \simeq& \alpha^{*}a_{y}+\alpha a_{y}^{\dagger } \\
S_{3}& \simeq& i\left(\alpha a_{y}^{\dagger }-\alpha^{*}a_{y}\right) \\
S_\theta&\simeq& \alpha^{*}  a_y e^{-i\theta}+\alpha a_y^{\dagger}e^{i\theta}
\end{eqnarray}
\end{subequations}
with
\begin{equation}\label{comutS} 
\left[ S_\theta,S_{ \theta^{\prime} }\right] \simeq 2i \vert \alpha \vert^{2}  sin\left( \theta^{\prime}-\theta\right)
\end{equation}

In particular,

\begin{equation}\label{comutapp} 
\left[ S_\theta,S_{ \theta+\frac{\pi}{2}}\right] \simeq 2i \vert \alpha \vert^{2} 
\end{equation}
which leads to the uncertainty relation:
\begin{equation}
\Delta S_{\theta}\Delta S_{( \theta+\frac{\pi}{2})} \geqslant  \vert \alpha \vert^{2}
\end{equation}

Introducing $X_{\theta}\equiv S_\theta/\vert \alpha\vert$ and $P_{\theta}\equiv S_{\theta+\frac{\pi}{2}}/\vert \alpha\vert$  we have:
\begin{equation}\label{comutxp} 
\left[ X_\theta,P_{\theta}\right] = 2i
\end{equation}
which is the usual commutation rule for orthogonal quadratures. In consequence these operators must obey the Heisenberg uncertainty relation:
\begin{equation}\label{Heisenberg}
\Delta X_{\theta}\Delta P_{\theta} \geqslant 1
\end{equation}

The observables $S_{\theta}$ and $S_{\theta+\frac{\pi}{2}}$ as well as $\vert \alpha \vert^{2}$ and therefore $X_\theta$ and $P_{\theta}$, are accessible to measurement. In this paper we consider a phase-space description of the field state in the $X,P$ (or $S_{\theta},S_{\theta+\frac{\pi}{2}}$) plane.


\begin{figure}[h]
\includegraphics[width=8cm]{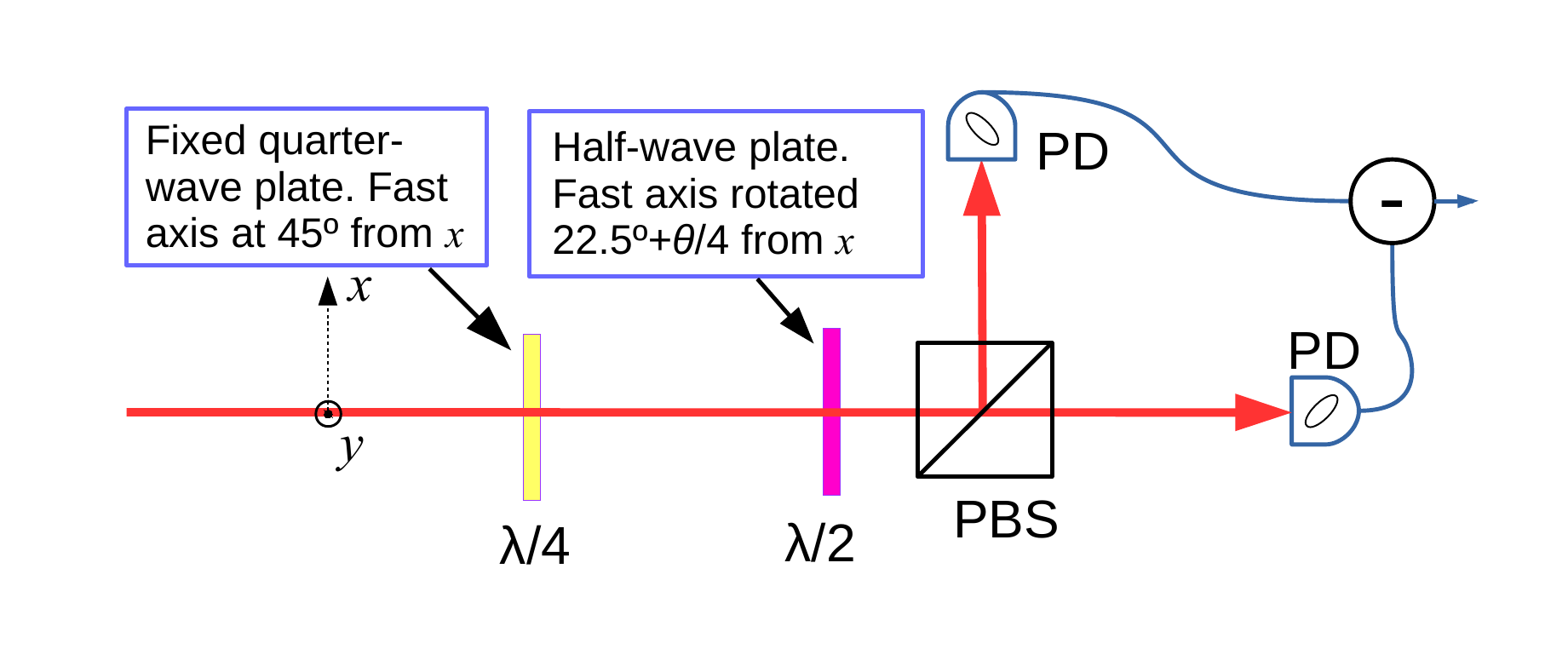}
\caption{(Color on-line) Experimental scheme for the measurement of the Stokes operator $S_\theta$. The angle $\theta$ is selected by the orientation of the half-wave plate axis. PBS: polarization beam-splitter. PD: photodetector.} \label{setup}
\end{figure}

The measurement of the generalized operator $S_\theta$ can be achieved using the arrangement shown in Fig. \ref{setup}. A fixed quarter-wave plate positioned with its principal axis oriented at $\pi/4$ with respect to the axis $x$ is followed by a half-wave plate whose fast axis forms an angle of $\pi/8+\theta/4$ with respect to $x$. With this arrangement, the two photodectectors outputs are respectively proportional to $n_{\theta +}$ and $n_{\theta -}$ i.e. the photon numbers corresponding to two orthogonal elliptical polarizations. In the case of our experiment, the axis of the two ellipsis are oriented at $\pm\pi/4$  with respect to the incident polarization. In consequence, the mean value of the two photodetectors outputs are balanced. This arrangement is simpler than the one that we have previously used \cite{BARREIRO11} which required a tilted waveplate and the calibration of the phase difference between orthogonal linear polarization as a function of the tilt angle. 

\subsection{Gaussian states}\label{gaussentanglement}

In recent years, quantum information processing with continuous variables has 
received large attention \cite{BRAUNSTEIN05}. A fundamental role is played by 
quantum systems prepared in Gaussian states \cite{WEEDBROOK12, FERRARO}.  
Such states have been used to demonstrate quantum correlations between two 
separate systems. Over the past decade, a significant amount of work has been 
devoted to the demonstration and measurement of entanglement using continuous 
variables 
\cite{Duan00,LAURAT05B,DAURIA09,YUKAWA08,HAGE08,YONEZAWA04,YONEZAWA07,DIGUGLIELMO07,SU12,AOKI03,UKAI11}. More recently 
\cite{MADSEN12,BLANDINO12,VOGL13,HOSSEINI14}, Gaussian states were used in the 
determination of the quantum discord between two systems. 


A quantum state describing a continuous variable is said to be Gaussian if its Wigner function defined over the $X,P$ plane is Gaussian. A two-dimensional Gaussian function is entirely defined by the covariance matrix:

\begin{equation}\label{covario} 
\left( \begin{matrix}
 \Delta X^2  &  \Delta XP \\
\Delta XP & \Delta P^2
\end{matrix}\right)
\end{equation}
where for any two operators $Y,Z$ we use $\Delta YZ\equiv \frac{1}{2}\left\langle YZ+ZY\right\rangle -\left\langle Y\right\rangle\left\langle Z\right\rangle$.

For a given $\theta$, it is easily shown that the covariance matrix is given by:

\begin{equation}\label{covadonga}
\left( \begin{matrix}
 \Delta X_{\theta}^2  & \frac{1}{2}\left[  \Delta X_{\theta+\frac{\pi}{4}}^2 - \Delta X_{\theta-\frac{\pi}{4}}^2  \right] \\
\frac{1}{2}\left[  \Delta X_{\theta+\frac{\pi}{4}}^2 -\Delta X_{\theta-\frac{\pi}{4}}^2 \right]& \Delta X_{\theta+\frac{\pi}{2}}^2 
\end{matrix}\right)
\end{equation}\\

All terms in (\ref{covadonga}) can be experimentally measured. Once the covariance matrix is known for a given choice of the angle $\theta$ it can be easily calculated for different angles through a rotation transform (see Eq. \ref{giro}). As the value of $\theta$ is varied, the variance $ \Delta X_{\theta}^2 $  describes an ellipsis. For the proper choice of $\theta$ corresponding to the orientation of one of the ellipsis main axis, the covariance matrix becomes diagonal. If $ \Delta X_{\theta}^2 <1$  (see Eq. \ref{Heisenberg}) the field is squeezed and the angle $\theta$ identifies the corresponding quadrature. \\


For two modes ($a$ and $b$) a Gaussian state is entirely characterized by the covariance matrix:
    
\begin{equation}\label{covarianza2} 
\left( \begin{matrix}
\Delta X_a^2  &  \Delta X_aP_a &  \Delta X_aX_b  &  \Delta X_aP_b \\
\Delta P_aX_a & \Delta P_a^2 &  \Delta P_aX_b  &  \Delta P_aP_b \\
\Delta X_bX_a  &  \Delta X_bP_a &  \Delta X_b^2  &  \Delta X_bP_b \\
\Delta P_bX_a & \Delta P_bP_a &  \Delta P_bX_b  &  \Delta P_b^2
\end{matrix}\right)
\end{equation}

All the coefficients in (\ref{covarianza2}) can be experimentally measured using the setup shown in Fig. \ref{setupexp}. The coefficients referring to a single mode are determined as in (\ref{covadonga}). The two-mode coefficients of the covariance matrix are of the form:
\begin{equation}\label{cruzados}
\Delta Y_aZ_b\equiv \dfrac{\Delta S_{\theta a}S_{\theta^{\prime} b}}{\vert \alpha_{a} \alpha_{b}\vert}
\end{equation} 
where $\vert \alpha_{a} \vert^{2}$ and $\vert \alpha_{b} \vert^{2}$ are given by the shot noise levels on modes $a$ and $b$ respectively.

Using the setup shown in Fig. \ref{setupexp}, the variances $\Delta(S_{\theta a}\pm S_{\theta^{\prime} b})^2$, corresponding to the sum and difference of the two balanced detectors outputs, can be measured. With the help of the identity, $\Delta S_{\theta a}S_{\theta^{\prime} b}=[\Delta(S_{\theta a}+ S_{\theta^{\prime} b})^2-\Delta(S_{\theta a}- S_{\theta^{\prime} b})^2]/4$ and the previously determined values of $\vert \alpha_{a} \vert$ and $\vert \alpha_{b} \vert$ the two-mode coefficients are readily computed.\\

Williamson's theorem \cite{WILLIAMSON36} ensures that the covariance matrix can be diagonalized through the application of a symplectic transformation. Also, the covariance matrix is characterized by four invariants under symplectic transformations. The two eigenvalues of the diagonal form are the (degenerate) symplectic eigenvalues $\nu_{+}$ and $\nu_{-}$ which can be expressed in terms of the symplectic invariants \cite{SERAFINI04}. The following properties derive from the symplectic invariants and eigenvalues.
\subsubsection{Consistency}
A covariance matrix must satisfy some constrains to represent a physical Gaussian state. In order to verify the Heisenberg uncertainty relation, the symplectic eigenvalues $\nu_k$ must verify \cite{ADESSO14}:
\begin{equation}\label{consistencia}
\nu_k \geq 1 
\end{equation}

\subsubsection{Entanglement}
The positivity of the partial transposed matrix is a necessary condition for separability \cite{SIMON00}. It is also a sufficient condition in the case of two-mode Gaussian states \cite{WERNER01}. If $\tilde{\nu}_{-}$ and $\tilde{\nu}_{+}$ with $(\tilde{\nu}_{+}>\tilde{\nu}_{-})$ are the symplectic eigenvalues of the partially transposed covariance matrix, the state is entangled if $\tilde{\nu}_{-} <1$. In addition, the amount of entanglement can be quantified through the logarithmic negativity $LN=-\sum_{k}log(\tilde\nu_k)$ (the sum extending over all $\tilde\nu_k < 1$) or $LN=0$ if $\tilde\nu_k \geq 1, \forall k$  \cite{VIDAL02}.
\subsubsection{Quantum Gaussian discord}
Entanglement does not account for all quantum correlations. In fact, quantum correlations can exist for separable mixed states. The quantum discord  \cite{OLLIVIER01} intends to quantify quantum correlations beyond entanglement. Initially defined for discrete quantum systems, its definition  was extended to Gaussian states \cite{GIORDA10,ADESSO10}. The optimality of the quantum Gaussian discord ($QGD$) is discussed in \cite{Pirandola14}. The $QGD$ can be directly computed from the four symplectic invariants of the covariance matrix \cite{ADESSO10}.

\section{Experiment}\label{experiment}

\begin{figure}[h]
\includegraphics[width=8cm]{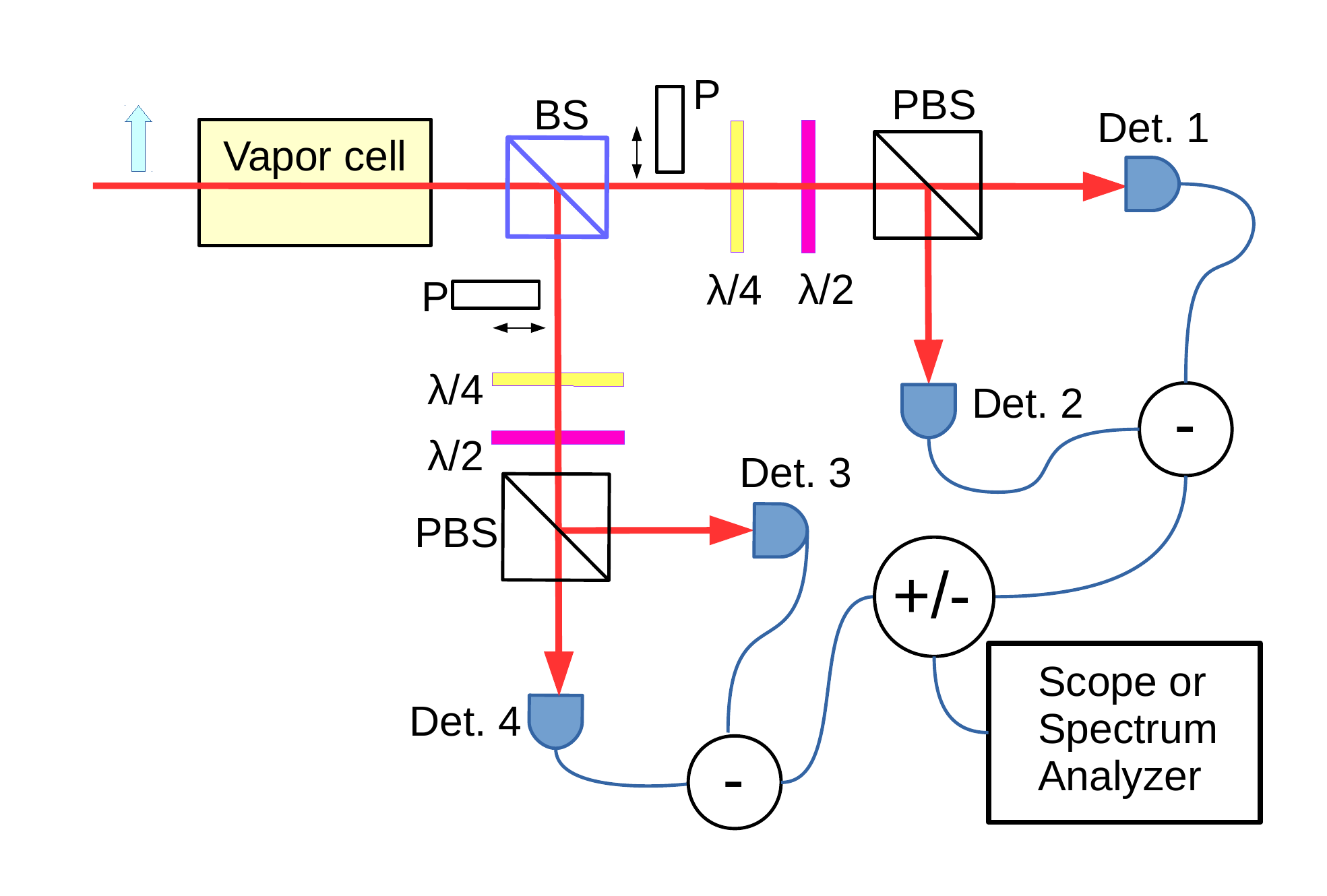}
\caption{(Color on-line) Experimental setup used for the characterization of a two-mode state corresponding to the output of the $\sim 50\%$ non-polarizing beam-splitter (BS). On each output a Stokes operator measurement setup is implemented. The balanced detector outputs can be either added or subtracted for global operator mesurements. } \label{setupexp}
\end{figure}

A scheme of the experimental setup is presented in Fig. \ref{setupexp}. The 
details of the laser source and the atomic cell environment have been previously 
described in \cite{BARREIRO11}. We remind the essential features. A 795 nm laser 
beam with approximately 40 mW of total power is linearly polarized and focussed 
with a 50 mm focal length at the center of a 5 cm long uncoated glass vapor 
cell containing isotopically pure $^{87}$Rb surrounded by a magnetic shield. The 
cell is heated to 50$^\circ$ C. The laser beam is re-collimated after the cell. 
The laser frequency can be scanned around the Rb D1 transitions.

In the first set of experiments  described below the beam splitter (BS) shown in Fig. \ref{setupexp} was not present thus directly realizing the detection arrangement presented in Fig. \ref{setup}. To determine the shot-noise level, a linear polarizer is introduced on the beam path before the wave-plates. As a consequence, the light polarization perpendicular to that of the incident field is blocked and replaced by vacuum. In this case, the balanced detector output variance corresponds to $\Delta S_\theta^{2}=\vert \alpha \vert^{2} $ which represents the shot-noise level for the fluctuations of the Stokes operators $S_{\theta}$. In subsequent measurements, with the polarizer removed, the shot-noise level is scaled in proportion to the light intensity to account for the small attenuation introduced by the polarizer. 

In a second set of experiments the \emph{non-polarizing} beamsplitter BS was introduced. An approximately $50\%$ transmission and reflection BS was used. Two separate balance detections were implemented to measure Stokes operators on both output ports of the beamsplitter. As a consequence of the BS being non-polarizing, the intense polarization component of the field is the same on the two outputs. The wave-plate angles are referred to this common direction.  The AC outputs of the two balanced detectors are either added or subtracted to access the fluctuations of global observables of the system. The total signal is sent to an oscilloscope for a time domain study of the fluctuations or to a spectrum analyser (SA) that records the noise power within a given bandwidth around a selected central frequency. The shot-noise levels corresponding the each of the light beams are independently recorded. 

For a given combination of the waveplates, the laser is scanned around the D1 transitions of $^{87}$Rb and the corresponding fluctuations data is recorded. Using the data for the four choices of  $\theta=-\pi/4,0,\pi/4,\pi/2$ the covariance matrix (Eq. \ref{covario}) is determined as a function of the laser detuning. For each detuning the covariance matrix can be brought into its diagonal form and the minimum and maximum noise variances computed as well as the corresponding quadrature angles. 

\section{Results}\label{results}
\subsection{Gaussianity test}\label{Gaussianity}

We have initially tested our assumption of Gaussian statistics of the light fluctuations. To assert the Gaussian nature of the field state is not a simple task \cite{PauloN}. We have approached this question empirically by examining the data corresponding to the balanced detector output as a function of time for different choices of the Stokes operator and the field. For each $S_{\theta}$, we have recorded 40 oscilloscope traces consisting of 2500 data points taken over a 250 $\mu$s time interval. Since the temporal record of the fluctuations is plagued by technical noise excess, specially at low frequency, we have numerically band-pass filtered the data string to extract the fluctuations corresponding to a specific frequency interval $[\Omega-{\Delta}/{2}, \Omega+{\Delta}/{2}]$ where $\Omega$ is the noise central frequency and $\Delta$ the bandwidth. The filtering procedure consisted in Fourier transforming the data string, replacing by zeros the points corresponding to frequencies whose \emph{absolute value}
 lies outside the frequency interval and performing the inverse Fourier transform. 

\begin{figure}[h]
\includegraphics[width=8cm]{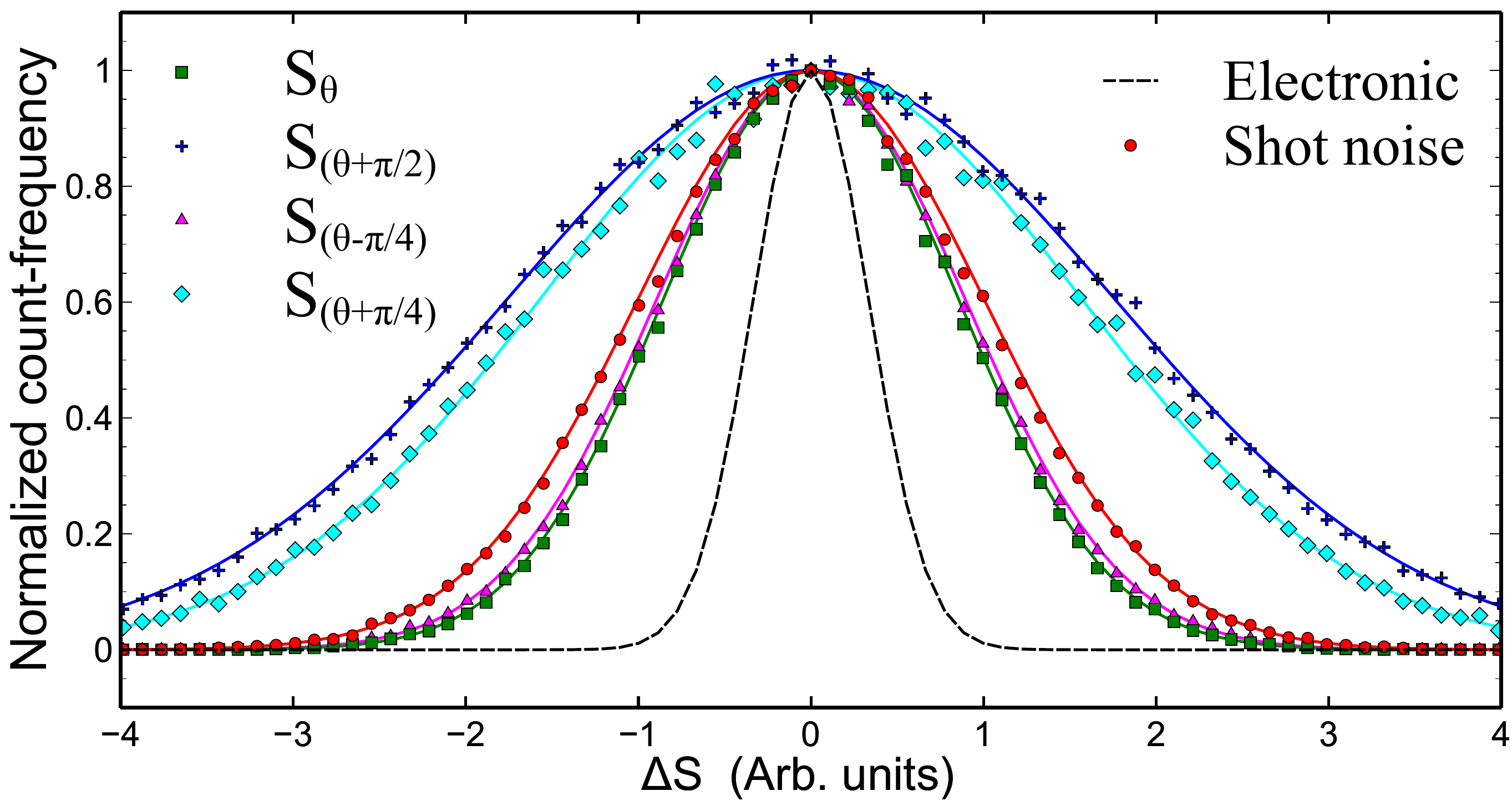}
\caption{(Color on-line) Symbols: Normalized histograms of the measurement of the filtered  ($\Omega=3$ MHz, $\Delta=400$ kHz) fluctuations of the Stokes operators. Solid lines: Gaussian functions with the same variance than the corresponding data. Dashed line: Residual electronic-noise level. The laser is tuned near the $^{87}$Rb $F_g=2 \rightarrow F_e=2$ transition at the position of maximum squeezing.} \label{gaussianas}
\end{figure}

The histogram of the filtered time domain fluctuations are presented in Fig. \ref{gaussianas} for different choices of the Stokes operator $S_{\theta}$. The histogram of the corresponding shot noise fluctuations is also shown. All histograms are well fitted by a normalized Gaussian function with the corresponding variance. The data in Fig. \ref{gaussianas} corresponds to a maximum squeezing of 1.6 dB with no correction for losses. 

\begin{figure}[h]
\includegraphics[width=8cm]{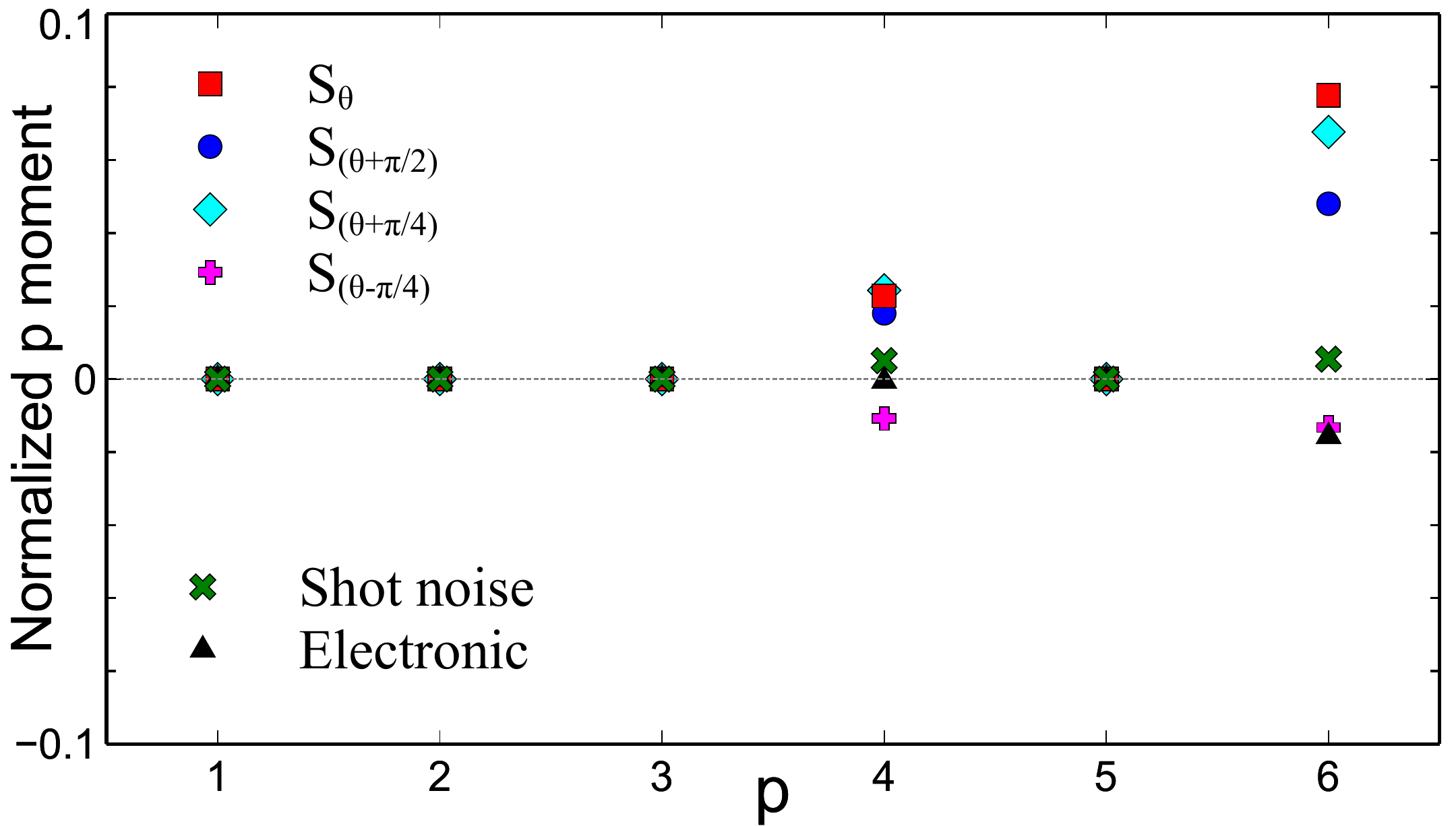}
\caption{(Color on-line) Normalized statistical moments for the measurements of Stokes operators. The normalized moment of order $p$ is defined as: ${\left\langle x^p \right\rangle }/\left[ {\left\langle x^2 \right\rangle^{\frac{p}{2}}(p-1)!!}\right] -\varepsilon_p$ where $\varepsilon_p=1$ if $p$ is even and zero otherwise. Here $x$ represents the result of a measurement. The normalized moments are zero for a Gaussian distribution.} \label{momentos}
\end{figure}

An additional test of the Gaussian statistics of the Stokes operators fluctuations results from the computation of the stochastic moments. For a Gaussian distribution, the odd moments are all zero and  for $p$ even they obey: $\left\langle x^p \right\rangle = \left\langle x^2 \right\rangle^{\frac{p}{2}}(p-1)!!$. Figure \ref{momentos} shows the comparison of the statistical moments of the recorded data to the corresponding value for a Gaussian distribution for moments up to $p=6$. The odd moments are zero within the measurement uncertainty and the even moments differ from the expecting Gaussian value in less than $8\%$. 

Supported on these observations, we conclude that the assumption of Gaussian nature of the field state is reasonable in the context of this work.\\ 

\subsection{Single Stokes operator}\label{single}

\begin{figure}[h]
\includegraphics[width=8cm]{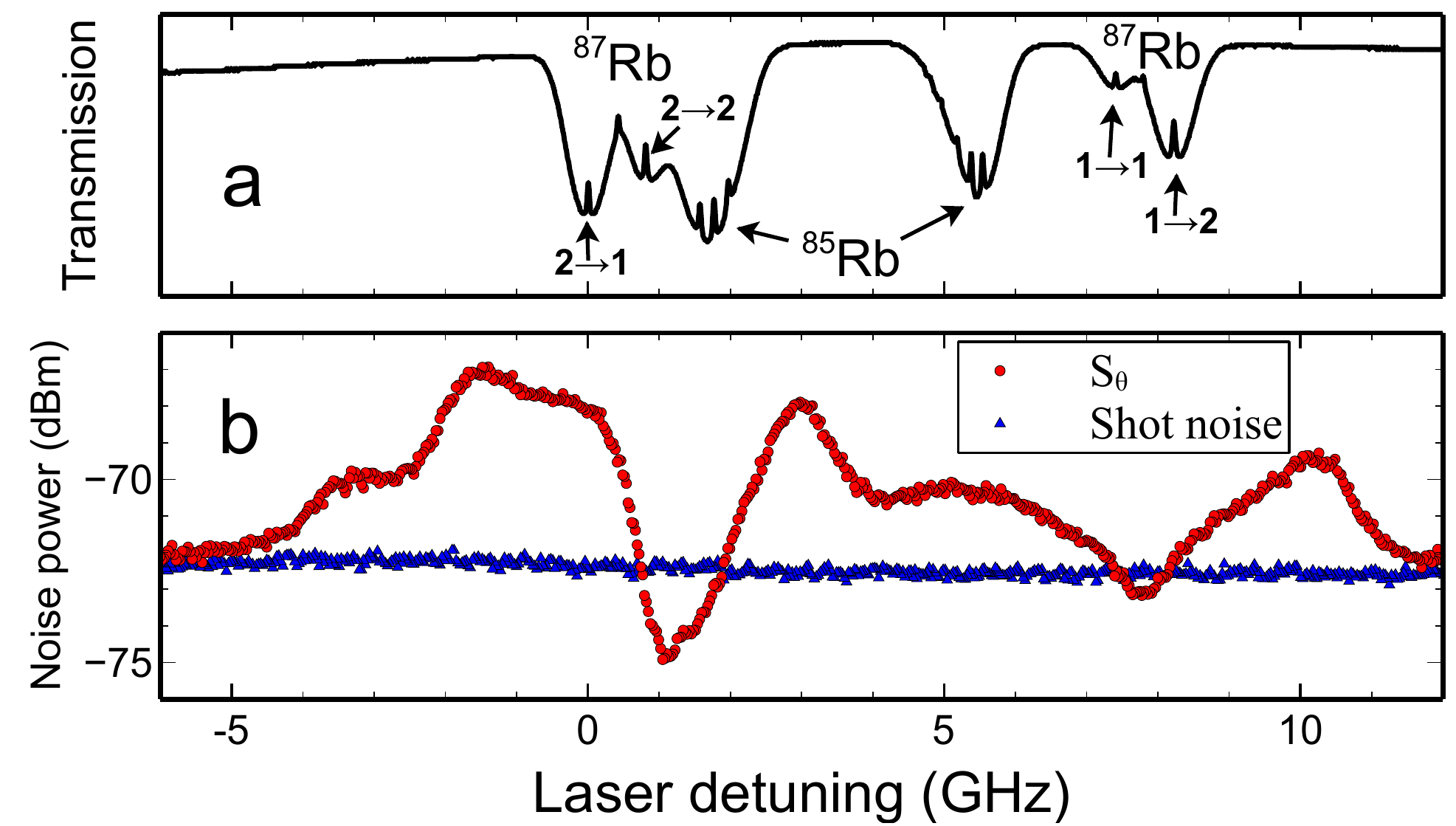}
\caption{(Color on-line) a) Reference-cell saturated-absorption signal used for frequency calibration. The hyperfine transitions of the $^{87}$Rb D1 line are indicated. b) Circles: Noise power at 2.7 MHz for the Stokes operator $S_{\theta}$ presenting maximum squeezing around the $F=2 \rightarrow F^{\prime}=2$  transition in a cell containing isotopically pure $^{87}$Rb. Triangles: Shot noise level.} \label{ancho}
\end{figure}

Figure \ref{ancho} presents the noise power (recorded with a spectrum analyser) 
around 2.7 MHz (resolution bandwidth 100 KHz) of a single Stokes operator as a 
function of laser detuning.  In Fig. \ref{ancho}, $S_{\theta}$ was chosen to 
correspond to the maximum squeezing occurring near the $F=2 \rightarrow 
F^{\prime}=2$ transition. Smaller squeezing is also observed around the $F=1 
\rightarrow F^{\prime}$ transitions. For other laser detunings the fluctuations 
of $S_{\theta}$ are above the shot-noise level. It is worth noticing the broad 
frequency range for which the light fluctuations are affected by the atomic 
transitions. It extends over a range of around 17 GHz, much larger than the 
total D1 absorption spectrum of rubidium. A slow decrease of the 
excess noise structure with the laser detuning from atomic resonances was 
predicted by the numerical simulations reported in \cite{MIKHAILOV09}.

\subsection{Phase-space description of one field}\label{phase}

\begin{figure}[h]
\includegraphics[width=8cm]{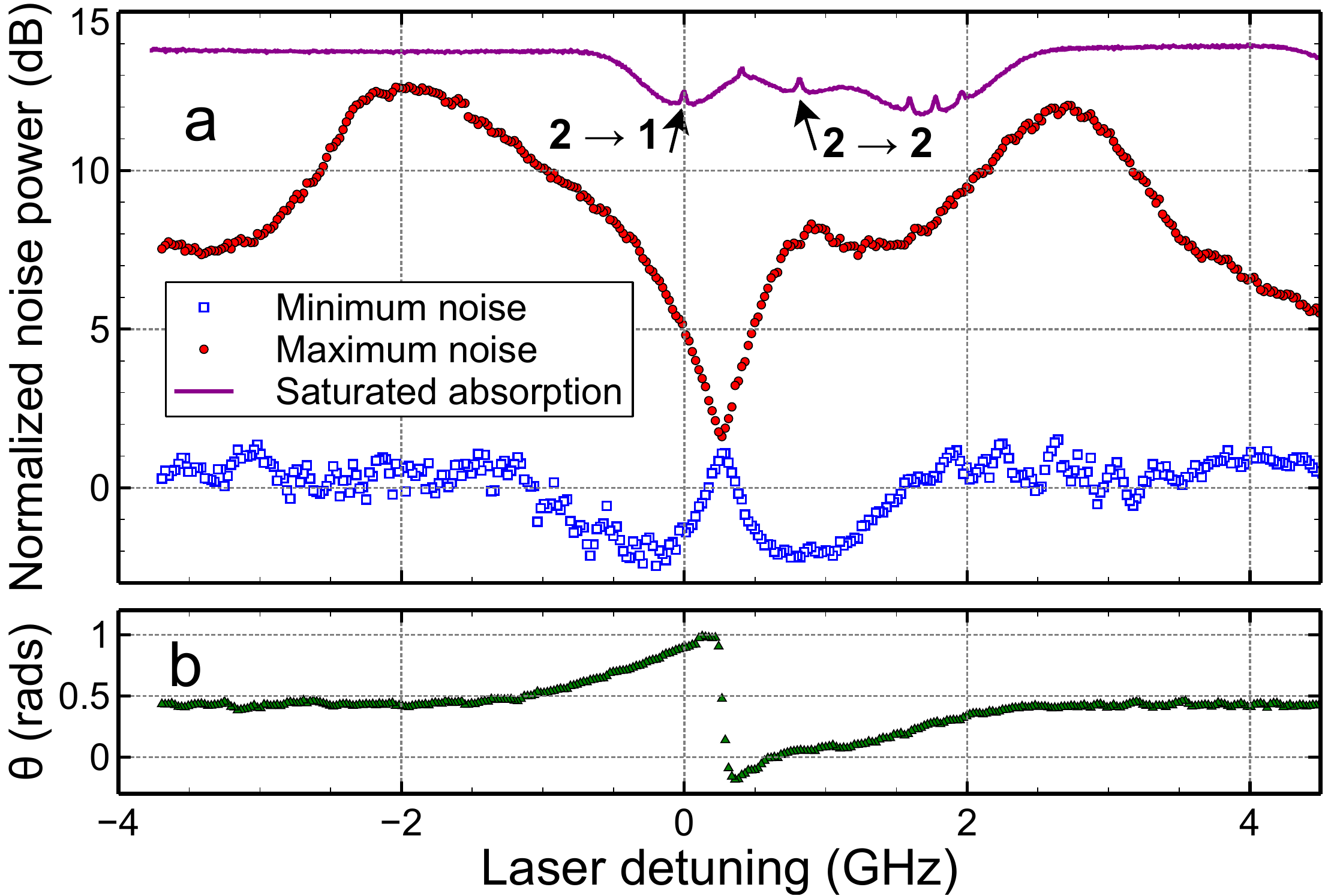}
\caption{(Color on-line) (a) Quadrature noise power normalized to the shot-noise level as a function of laser detuning. Squares: Minimum quadrature noise. Circles: Maximum quadrature noise. Solid: Reference saturated absorption signal with relevant $^{87}$Rb D1 transitions indicated. (b) Quadrature angle corresponding to the minimum noise. (See also  \cite{Comment1}).} \label{doscuadraturas}
\end{figure}

In order to have access to the covariance matrix, we have recorded with a 
spectrum analyser (central frequency 2.7 MHz, resolution bandwidth 100 kHz) the 
fluctuations of the four Stokes operators 
$S_{\theta-\pi/4},S_{\theta},S_{\theta+\pi/4},S_{\theta+\pi/2}$. For convenience 
 we chose $S_{\theta}$ to correspond to $S_3$. The covariance matrix was 
constructed according to (\ref{covadonga}) and diagonalized to identify the 
minimum and maximum noise variances and the corresponding quadrature angle. The 
results are presented in Fig. \ref{doscuadraturas} (see also \cite{Comment1}). Similar 
spectra of the minimum and maximum noise variances were presented in 
\cite{MIKHAILOV08,MIKHAILOV09,AGHA10}. Notice the considerable simplification 
arising from the combination of Stokes operators detection and Gaussian 
analysis. In \cite{MIKHAILOV08,MIKHAILOV09}, a sophisticated active 
stabilization of the phase of the local oscillator was implemented in order to 
track the field quadrature corresponding to maximum or minimum noise. In 
\cite{AGHA10} this was achieved manually for a discrete number of laser 
detunings.  Also these articles did not report the determination of the quadrature 
angle corresponding to squeezing possibly due to difficulties in the local 
oscillator phase calibration in the interferometric setup. By contrast, in our 
setup there are no interferometric instabilities affecting the choice of 
$S_{\theta}$, no stabilization loop is required since the data is recorded for a 
fixed value of the angle $\theta$ and the determination of the noise variances 
and corresponding quadrature angles readily arise from the numerical covariance 
matrix diagonalization.

The results presented in Fig. \ref{doscuadraturas} reveal interesting  features about the noise structure. As previously observed \cite{MIKHAILOV08,MIKHAILOV09,AGHA10,BARREIRO11}, squeezing is present around the $F=2 \rightarrow F^{\prime}=1$ and $F=2 \rightarrow F^{\prime}=2$ albeit for different quadrature angles. Interestingly enough, for the laser detunings where squeezing does not occur, the minimum noise variance remains close to the shot-noise level. On the other hand the maximum noise variance presents excess noise above the limit imposed by the Heisenberg uncertainty relation. 

Perhaps the more interesting feature in Fig. \ref{doscuadraturas} is the existence of a laser frequency between the two hyperfine transitions for which the maximum and minimum noise variances are nearly equal and very close to the shot-noise level. The existence of this singular point in the noise spectrum can be related to the cancellation of the PSR effect around this position \cite{MATSKO02}. It is interesting to notice that not only the squeezing is suppressed but very little excess noise is added by the atomic interaction. The squeezing occurs on either side of such noise cancellation point. 

Our method readily allows the observation of the rotation of the Stokes operator quadrature noise ellipsis as a function of laser detuning \cite{Comment1}. A steep variation of the orientation of the noise ellipsis over 72 degrees occurs around the noise cancellation point. It was recently suggested \cite{Kimble01,Horrom13} that the control of the orientation of squeezed light noise ellipsis could be applied to precision improvement in gravitational wave detection interferometry. Rotation of the noise ellipsis was recently observed in squeezed light generated via four-wave-mixing in an atomic system  \cite{Corzo13}.
 
\subsection{Two-field state characterization}\label{tf}

\begin{figure}[h]
\includegraphics[width=8cm]{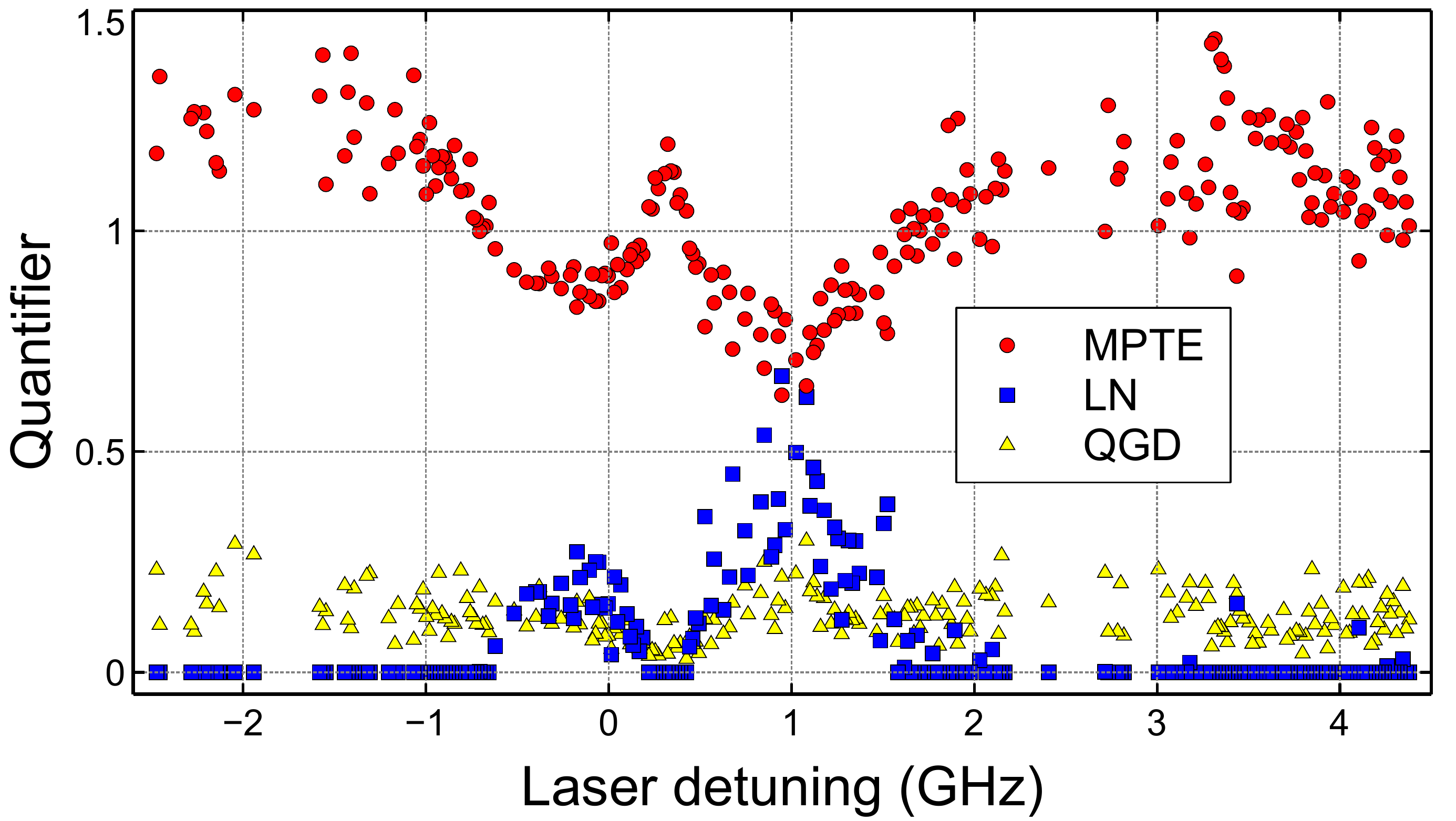}
\caption{(Color on-line) Two-mode state quantifiers for the two beamsplitter outputs as a function of laser detuning. Circles: minimum partial transpose eigenvalue  ($MPTE\equiv\tilde{\nu}_{-}$). Squares: logarithmic negativity ($LN$). Triangles: quantum Gaussian discord ($QGD$). } \label{doscampos}
\end{figure}

We have recorded the covariance matrix of the compound system for each value of the laser detuning following the procedure described in Section \ref{gaussentanglement}. 

Since the experimentally determined covariance matrix is affected by noise, we have tested each covariance matrix for consistency and physical significance. We have discarded all the covariance matrices that did not satisfy the physical consistency requirement (Eq. \ref{consistencia}).  The data complying to this requirement was used for symplectic invariants determination and computation of the symplectic eigenvalue $\equiv\tilde{\nu}_{-}$, the logarithm negativity and the quantum Gaussian discord between modes $a$ and $b$ conditioned to measurements on $b$ \citep{ADESSO10}. 

The results are presented in Fig. \ref{doscampos}. In spite of the significant amount of noise, the positivity of the partial transpose criterion and the logarithmic negativity indicate entanglement near the laser frequency range for which squeezing is present on single field Stokes operators (see Fig. \ref{doscuadraturas}). Notice that nonzero quantum Gaussian discord occur in all the laser frequency range shown in Fig. \ref{doscampos} indicating that the two-mode state presents some amount of nonclassical correlations even when it is separable. 

As a final remark, it is worth mentioning that the experimental procedure described above is independent of the intensity balance of the two outputs of the BS. Such balance will nevertheless have an influence on the entanglement and quantum Gaussian discord observed.

\section{Conclusions}
We have shown that the description of the field in terms of polarization (Stokes) states in combination with the assumption of Gaussian statistics results in considerable  simplification of the study of the quantum state of light transmitted through an atomic sample: The experimental setup is dramatically reduced. All interferometric instabilities are removed. The selection of the detected phase space quadrature corresponds to the rotation angle of a single waveplate. Finally, the complete state (noise ellipsis) reconstruction is computed through matrix diagonalization. 

Taking advantage of this simplicity, we have presented, for the first time to our knowledge, the complete evolution of the noise ellipsis describing light fluctuations around a given central frequency as a function of laser detuning. In the case of two light modes, obtained by parting a squeezed light beam in a beamsplitter, the use of Gaussian states formalism in combination with Stokes operator observations, allowed us to demonstrate entanglement and measure the quantum Gaussian discord indicating that the generated light states could be suitable for quantum information processing. 

\section{\label{Acknoledgments}Acknoledgments}
We wish to thank P. Nussenzveig for useful suggestions. This work was supported by CSIC and ANII.

\end{document}